\documentclass[smallextended,epjc3]{svjour3}          % twocolumn

\usepackage{lineno}
\usepackage{siunitx}
\usepackage{amssymb}
\usepackage{amsmath}

\RequirePackage[T1]{fontenc}

\smartqed  % flush right qed marks, e.g. at end of proof

\RequirePackage{graphicx}
\RequirePackage{mathptmx}      % use Times fonts if available on your TeX system
\RequirePackage{flushend}
\RequirePackage[numbers,sort&compress]{natbib}
\RequirePackage[colorlinks,citecolor=blue,urlcolor=blue,linkcolor=blue]{hyperref}

\journalname{Eur. Phys. J. C}

\begin{document}
%\linenumbers

\title{Dependence of atmospheric muon flux on seawater depth measured with the first KM3NeT detection units}
\subtitle{The KM3NeT Collaboration}

% ----- Start automatically generated KM3NeT info
% ----- Start author list
\author{
M.~Ageron\thanksref{a}\and 
S.~Aiello\thanksref{b}\and 
F.~Ameli\thanksref{c}\and 
M.~Andre\thanksref{d}\and 
G.~Androulakis\thanksref{e}\and 
M.~Anghinolfi\thanksref{f}\and 
G.~Anton\thanksref{g}\and 
M.~Ardid\thanksref{h}\and 
J.~Aublin\thanksref{i}\and 
C.~Bagatelas\thanksref{e}
% ...
}
% ----- End author list
% ----- Start affiliation list
\institute{
\label{a} Aix~Marseille~Univ,~CNRS/IN2P3,~CPPM,~Marseille,~France
\and
\label{b} INFN, Sezione di Catania, Via Santa Sofia 64, Catania, 95123 Italy
\and
\label{c} INFN, Sezione di Roma, Piazzale Aldo Moro 2, Roma, 00185 Italy
\and
\label{d} Universitat Polit{\`e}cnica de Catalunya, Laboratori d'Aplicacions Bioac{\'u}stiques, Centre Tecnol{\`o}gic de Vilanova i la Geltr{\'u}, Avda. Rambla Exposici{\'o}, s/n, Vilanova i la Geltr{\'u}, 08800 Spain
\and
\label{e} NCSR Demokritos, Institute of Nuclear and Particle Physics, Ag. Paraskevi Attikis, Athens, 15310 Greece
\and
\label{f} INFN, Sezione di Genova, Via Dodecaneso 33, Genova, 16146 Italy
\and
\label{g} Friedrich-Alexander-Universit{\"a}t Erlangen-N{\"u}rnberg, Erlangen Centre for Astroparticle Physics, Erwin-Rommel-Stra{\ss}e 1, 91058 Erlangen, Germany
\and
\label{h} Universitat Polit{\`e}cnica de Val{\`e}ncia, Instituto de Investigaci{\'o}n para la Gesti{\'o}n Integrada de las Zonas Costeras, C/ Paranimf, 1, Gandia, 46730 Spain
\and
\label{i} APC, Universit{\'e} Paris Diderot, CNRS/IN2P3, CEA/IRFU, Observatoire de Paris, Sorbonne Paris Cit\'e, 75205 Paris, France
% ...
}
% ----- End affiliation list
% ----- End automatically generated KM3NeT info

\thankstext{corr1}{Corresponding author: simone.biagi@infn.it}
\thankstext{corr2}{Corresponding author: lincetto@cppm.in2p3.fr}
\thankstext{corr3}{Corresponding author: kmelis@nikhef.nl}

\date{Received:  / Accepted:}

\maketitle

\begin{abstract}
KM3NeT is a research infrastructure located in the Mediterranean Sea, that will consist of two deep-sea Cherenkov neutrino detectors. With one detector (ARCA), the KM3NeT Collaboration aims at identifying and studying TeV-PeV astrophysical neutrino sources. With the other detector (ORCA), the neutrino mass ordering will be determined by studying GeV-scale atmospheric neutrino oscillations.
The first KM3NeT detection units were deployed at the Italian and French sites between 2015 and 2017. In this paper, a description of the detector is presented, together with a summary of the procedures used to calibrate the detector in-situ. Finally, the measurement of the atmospheric muon flux between $2232-$\SI{3386}{m} seawater depth is obtained.
\end{abstract}

\section{Introduction} 
\label{s:intro}

The KM3NeT (km$^3$-scale neutrino telescope) Collaboration is establishing a research facility in the Mediterranean Sea that will host a network of neutrino detectors \cite{loi}. 
Neutrinos represent an alternative to photons and cosmic rays to explore the high-energy Universe. They can emerge from dense objects and travel large distances, without being deflected by magnetic fields or interacting with radiation and matter.
In addition, the neutrino mass ordering can be determined using neutrinos produced in the earth atmosphere through interactions of cosmic rays \cite{loi}.

The ARCA (Astroparticle Research with Cosmics in the Abyss) detector is being installed at the KM3NeT-It site, \SI{80}{km} offshore the Sicilian coast in front of Capo Passero (Italy) at a sea bottom depth of about \SI{3450}{m}. 
The main physics goal of ARCA is the discovery and subsequent observation of astrophysical neutrino sources in the Universe \cite{arca}. About \SI{1}{km^3} of seawater will be instrumented with $\sim$130,000 photomultiplier tubes (PMTs) for the detection of Cherenkov light induced by charged particles produced in neutrino interactions. The geometry of ARCA is optimised to maximise its detection efficiency in the neutrino energy range \SI{1}{TeV}--\SI{10}{PeV}.

In parallel, the KM3NeT Collaboration started the construction of the ORCA (Oscillation Research with Cosmics in the Abyss) detector at the KM3NeT-Fr site, \SI{40}{km} offshore Toulon (France) at a sea bottom depth of  about \SI{2450}{m}. ORCA shares the same technology and detector elements as ARCA, but in a denser configuration to detect neutrinos in the range 1--\SI{100}{GeV}. 
ORCA instruments  a volume of about \SI{8}{Mton}. The main physics goal is the determination of the neutrino mass ordering by measuring the oscillation probabilities of atmospheric neutrinos \cite{loi,orca}.

In this paper, a procedure to calibrate the PMT detection efficiencies using the light emitted by ${}^{40}$K decays is discussed. The flux of atmospheric muons produced in cosmic ray air showers forms a background to the primary objective of KM3NeT. A measurement of the depth dependence of the atmospheric muon flux will be presented. This measurement is used to verify the calibration procedure. The results are obtained with the first two detection units of ARCA (detector configuration referred to as ARCA2) and the first one of ORCA (ORCA1).

In Section~\ref{s:detector}, an overview of the ARCA and ORCA detectors and their detection elements is given. 
Section~\ref{s:calib} outlines the time and photon detection efficiency calibration procedures.
Section~\ref{s:sample} presents the data samples used in the analysis. 
The algorithms used to simulate the detector behaviour are described in Section~\ref{s:mc}.
The measurement of the atmospheric muon flux depth dependence is reported and discussed in Section~\ref{s:muons}.
Finally, the conclusions are given in Section~\ref{s:conclusion}.

\section{The KM3NeT detectors} 
\label{s:detector}

\begin{figure*}
\centering
\includegraphics[width=0.95\textwidth]{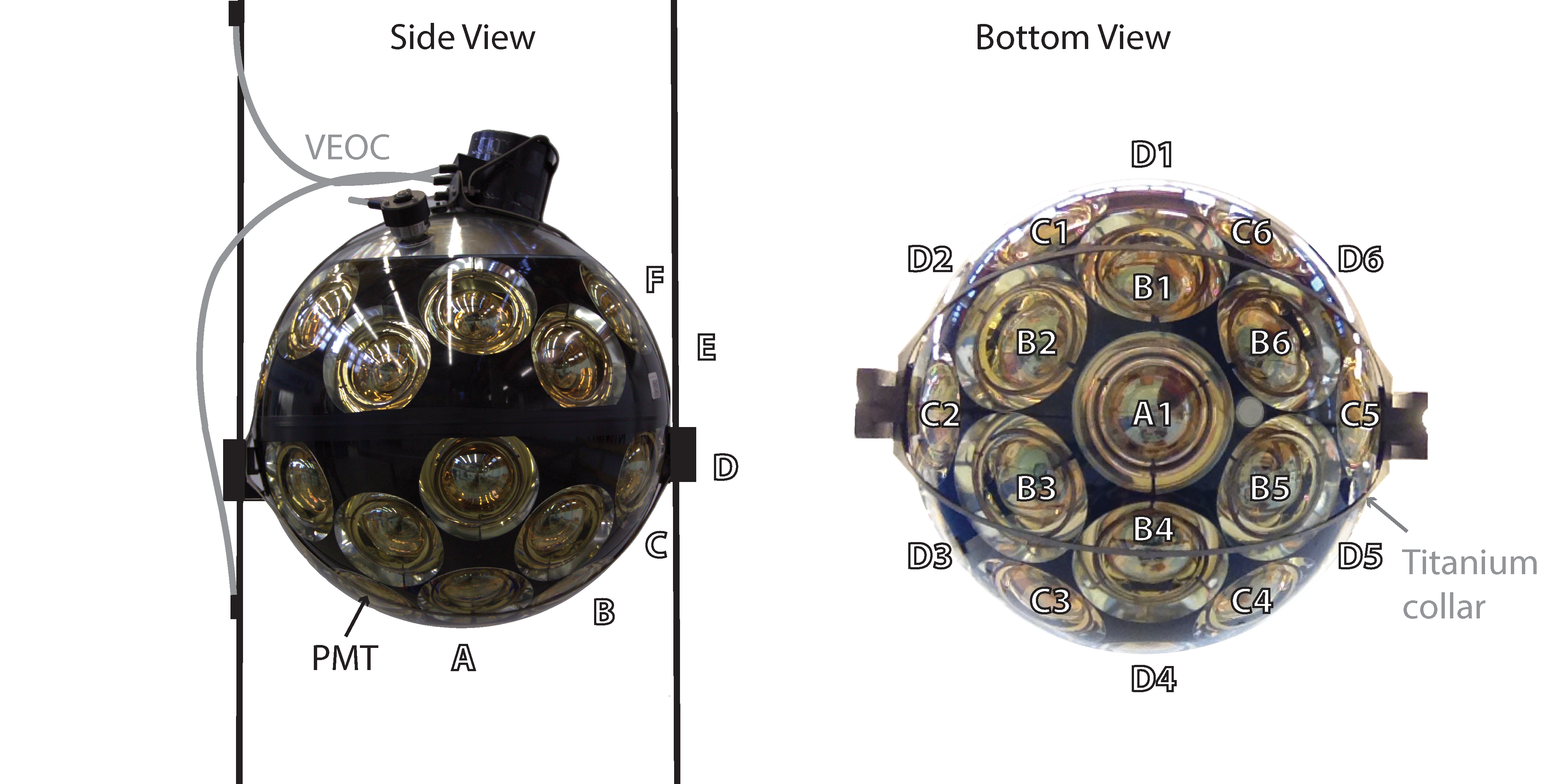}
\caption{Side view (left) and bottom view (right) of a KM3NeT DOM. See text for explanation of highlighted (mechanical) parts. The naming convention of the PMTs in rings and numbers is indicated.}
\label{f:DOMpicture}
\end{figure*}

ARCA and ORCA consist of three dimensional arrays of Digital Optical Modules (DOMs), installed in the deep waters of the Mediterranean Sea. The DOM is a pressure-resistant glass sphere housing 31 PMTs of \SI{80}{mm} diameter each (see Figure~\ref{f:DOMpicture}). These PMTs detect the Cherenkov light induced by relativistic charged particles propagating through the seawater. A transparent gel is employed to guarantee the optical coupling between the PMTs and the internal surface of the glass spheres. A reflector ring is mounted around each PMT to enhance its sensitive area \cite{expansion_cone}.

Several instruments are mounted inside the DOM to measure its position and orientation (compass, tilt-meter, acoustic sensor) and to time-calibrate the DOMs (light emitters based on LEDs). The real-time positioning of the DOMs is necessary as the detector elements can move under the influence of sea currents. A description of the DOM is given in Refs. \cite{ICRC_DOM_Bruijn,ppm-dom}.

A KM3NeT detection unit (DU) is a series of 18 DOMs connected by an electro-optical cable and arranged in a flexible and slender vertical string.
The glass sphere of each DOM is encircled by a titanium collar which is attached to two \SI{4}{mm} diameter ropes (Figure~\ref{f:DOMpicture}). 
Each DU is anchored to the seabed using a dead weight. A buoy is connected at the top of the DU to keep it close to vertical.

A set of 115 detection units constitutes a building block of the KM3NeT detectors. ARCA and ORCA differ in the granularity of the sensor modules. 
For the ARCA-type detection unit the vertical distance between the DOMs is \SI{36}{m}, resulting in a total height of the structure of \SI{700}{m}. The installation of two ARCA building blocks at the KM3NeT-It site is planned, with a horizontal distance between the DUs  of about \SI{90}{m} and a total instrumented volume of about \SI{1}{km^3}. 
In the detection units of ORCA, the DOMs are mounted at an average vertical distance of about \SI{9}{m} from each other, and the total height of the unit is \SI{250}{m}. ORCA plans the construction of one building block with a distance between the detection units of about \SI{20}{m}.

The analog signals from the PMTs are digitised using a custom electronic board mounted inside each DOM \cite{clb}. When one or more photons impinge on the PMT photocathode and the anode electrical signal crosses the threshold of a discriminator, a level zero (L0) hit is recorded. The crossing time and the time-over-threshold (ToT) of the waveform are recorded. The ToT gives a measure of the pulse amplitude. The high voltage setting of each PMT is tuned in-situ to give an average ToT of \SI{26.4}{ns} for a single photoelectron (p.e.). A default value corresponding to 0.3\,p.e. is set for the discriminator threshold. 

With an \emph{all-data-to-shore} approach \cite{antares_elec}, no data reduction is applied underwater. All L0 hits are assembled in time windows with a fixed size of \SI{100}{ms}, called \emph{L0 timeslices}. The timeslices are sent to a computer farm at the shore station via a network of submarine cables and junction boxes. 

The DU base contains the power and control electronics and the components of the optical network for the communication between the DOMs and the onshore station. It consists of a titanium cylinder attached to the anchor.
For power delivery and data transmission, a Vertical Electro-Optical Cable (VEOC) is attached to each DOM. The VEOC is a pressure-balanced, oil-filled, plastic tube that contains two copper wires for the power delivery (\SI{400}{V} DC) and 18 optical fibres for the data transmission. 
Two power conductors and a single fibre are branched out at the level of each DOM via a so-called breakout box. The breakout box contains a DC/DC converter (\SI{400}{V} to \SI{12}{V}) to supply each DOM with a suitable voltage.
The power conductors and optical fibre enter the glass sphere through a pressure-resistant penetrator. 

Onshore, once the time calibration is applied, dedicated trigger algorithms look in parallel for the physics events to be saved for offline analysis. A detailed discussion of trigger algorithms is reported in Ref. \cite{loi}. A level one (L1) coincidence is composed of two or more L0 hits on the same DOM within a time window of \SI{25}{ns}. Timeslices of L1 coincidences are stored for calibration and monitoring. In this paper, the L1 timeslices data are used. 

The data acquisition system must be able to handle the large data throughput produced by the detector \cite{ICRC_DAQ}. The average observed hit rate, dominated by the decay of $^{40}$K isotopes in the seawater, is about \SI{7}{kHz} per PMT, corresponding to about \SI{13}{Mbps} of data per DOM. Occasionally, the rate can be  higher due to bioluminescence activity. Bioluminescence is characterised by an increase (up to the MHz range) of the hit rate, lasting up to several seconds. The readout system of the DOM disables the channel acquisition as soon as the rate is estimated to be over \SI{20}{kHz} (\emph{high rate veto}).

\begin{figure}
\centering
\includegraphics[width=0.5\textwidth]{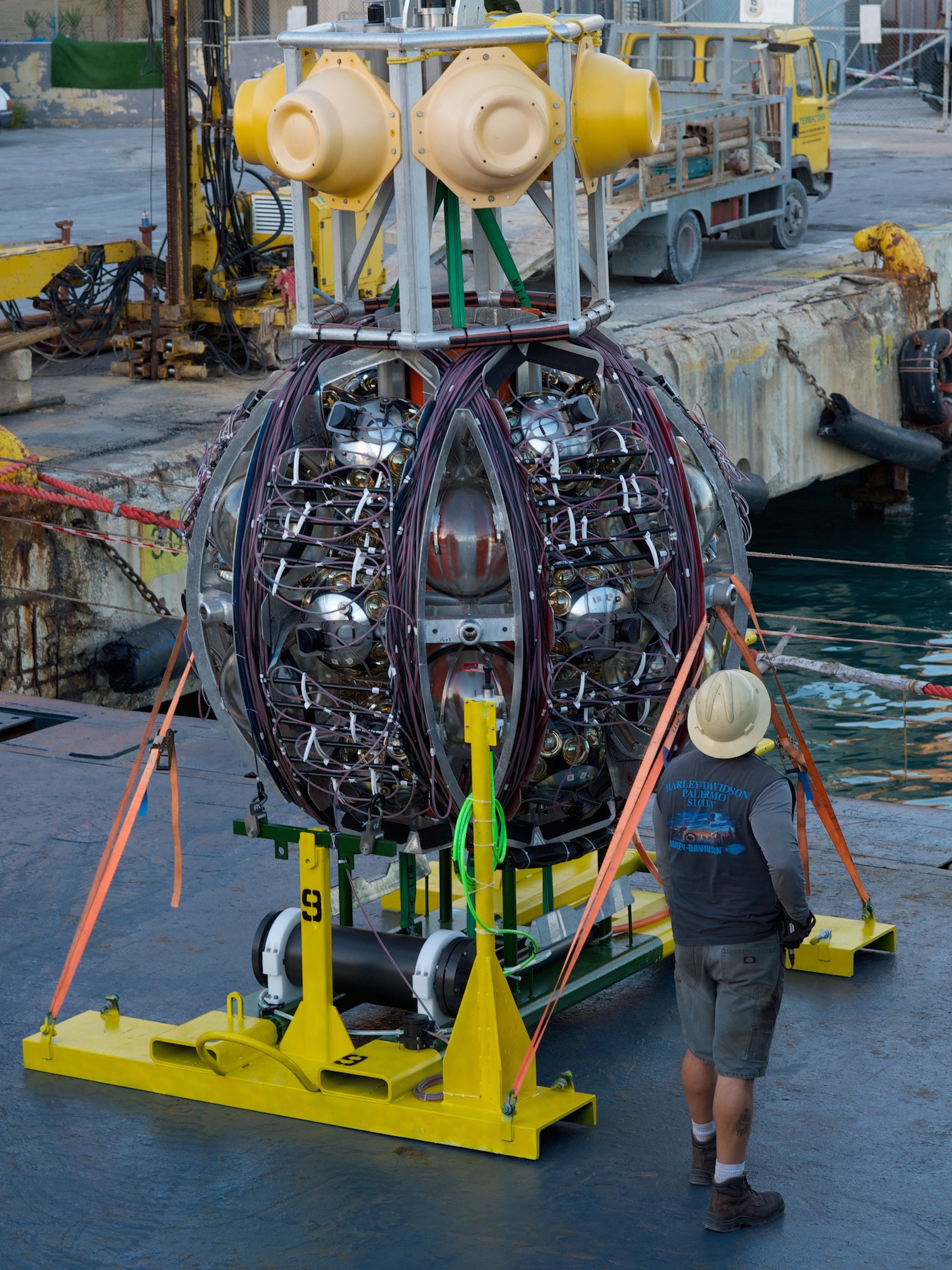} \ \ \ \ \ 
\caption{A detection unit  arranged on the launcher vehicle, ready to be deployed during a sea campaign. The yellow buoyancy system is visible on top of it.}
\label{f:deployment}
\end{figure}

For the deployment, the DU is furled onto a spherical structure with a radius of approximately \SI{2}{m}, called the Launcher of Optical Modules (LOM) \cite{LOM}, as can be seen in Figure~\ref{f:deployment}. 
Once at the seabed, the DU is connected to the seafloor network using a Remotely Operated Vehicle (ROV). After confirming the functionality of the DOMs, the LOM is remotely released and floats up to the surface, progressively unfurling the detection unit.

The first DU of the ARCA detector was deployed in December 2015, followed by two DUs in May 2016. The vertical alignment of the DUs was confirmed by visual inspection using the ROV. Data taking started immediately after deployment. One of the units deployed in May 2016 was recovered for inspection in July 2016. Due to electrical problems in the network infrastructure at the seabed, the operations at the KM3NeT-It site were on hold between April 2017 and January 2019, after which data taking continued. The DU of ORCA1 was installed in September 2017 and operated until mid-December 2017, when a failure of the main electro-optical cable occurred. The data taking was resumed in February 2019 following the replacement of a part in the main cable.

\section{Time and efficiency calibration}
\label{s:calib}

The majority of light  observed by the PMTs originates from the decays of the radioactive $^{40}$K isotope naturally abundant in sea salt. The Cherenkov light produced by the secondary electrons of the decay represents the main optical background for deep-sea neutrino detectors, but it is removed by requiring space-time correlations between DOMs. On the other hand, the $^{40}$K signature is particularly useful for  in-situ  calibration  of the PMTs contained in a single DOM  (intra-DOM  calibration) and  for long-term monitoring of the performance of the detector. 

The distribution of the time differences of L0 hits seen in coincidence by two PMTs $i,j$ is used to determine the relative time offset ($t0$), the transit time spread (TTS) and the photon detection efficiency ($\epsilon$) of the two PMTs involved \cite{ICRC_calib}. 
Here the time offset of each PMT in a DOM compensates for the (mean) transit time and differences in propagation time of the electrical signal of each PMT. Only the differences between the time offsets of the PMTs within a DOM can be fitted using this procedure. %The average time offset of the PMTs in a DOM is set to zero. 
In the fit, the sum of the PMT time offsets is constrained to zero.
The PMT photon detection efficiency reflects the combined effect of the PMT quantum efficiency, the collection efficiency, the angular acceptance of the PMT and absorption in the gel and the glass. 
In a DOM with $N=31$ PMTs, all the unique PMT pairs are taken into account, which corresponds to $N(N-1)/2=465$ pairs. 
Random coincidences of uncorrelated hits give an offset of the observed coincidence rate, independent of the hit time difference. This contribution is estimated from the observed L0 rates, and checked to be compatible with a fit to the tails of the distribution.
After subtracting the random  coincidence rate, the $465$ coincidence distributions of all PMT pairs within a DOM are simultaneously fitted with a Gaussian with mean $t0_i - t0_j$, variance $(TTS_i)^2 + (TTS_j)^2$ and area $f(\theta_{i,j}) \cdot \epsilon_i \cdot \epsilon_j$. The function $f(\theta_{i,j})$ is the expected coincidence rate as a function of the space angle  $\theta_{i,j}$ between the two PMTs. This parameterisation follows from dedicated GEANT4 \cite{Agostinelli:2002hh} simulations using a nominal DOM model \cite{ICRC_kulikovsky}. The function used is an updated version of the parameterisation presented in Ref. \cite{ICRC_calib}.

\begin{figure}
\centering
\includegraphics[width=0.7\textwidth]{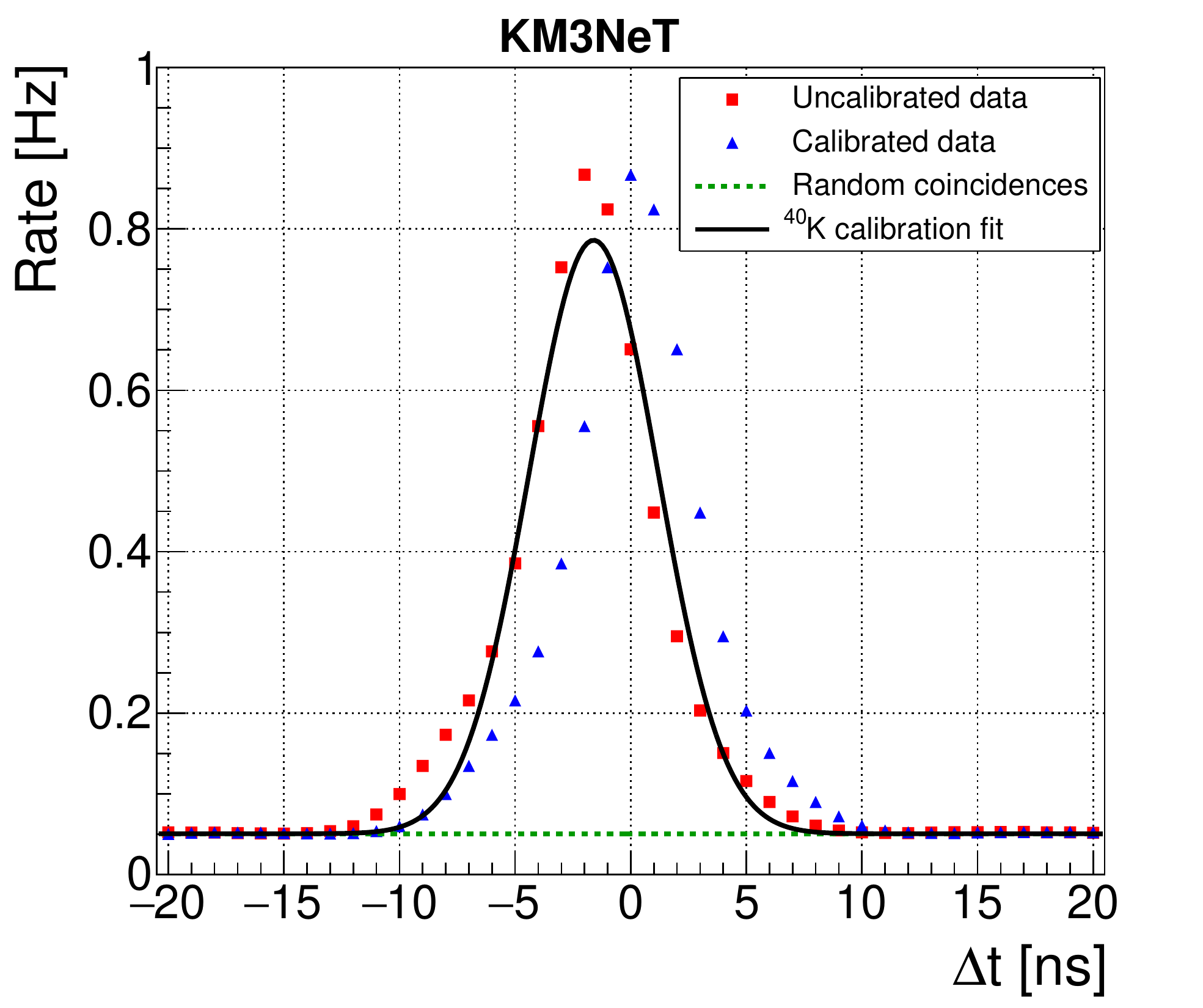}
\caption{Distribution of time differences between hits in coincidence for one typical pair of adjacent PMTs, before (red) and after (blue) calibration. In these, only statistical errors are shown, which are smaller than the symbols. A peak due to genuine $^{40}$K coincidences is observed above a flat background (green). The black line is a Gaussian fit of the calibration model (see text) to the uncalibrated data of all PMT pairs.}
\label{f:double_coinc}
\end{figure}

In Figure \ref{f:double_coinc}, the time difference distribution of one pair of PMTs (in this case two adjacent PMTs) is shown. The data before calibration (red) and after calibration (blue) are reported. The fitted random coincidence rate is shown in green. The result of the fit to all $465$ PMT pairs is shown in black. The difference between the measured distribution and the Gaussian fit can be explained as the result of non-Gaussian transit time distribution of the PMTs.
This effect has no impact on the peak of the distribution, i.e. on the determination of time offsets. Across all pairs, the average difference between the area of the Gaussian and the actual integral is 3\%, taken into account as a systematic error in the determination of the PMT efficiency.
The calibrated hit time difference distribution, shifted according to the fitted time offsets from the procedure outlined above, is shown in blue. As expected, the mean of the calibrated time difference distribution is zero.

The  obtained intra-DOM time offsets are used  to calibrate data  in the offline analysis, while the PMT efficiencies are used in simulations (see Section \ref{s:mc}). The time calibration between DOMs (inter-DOM calibration) is done with calibrated laser pulses before DU deployment. In-situ, the inter-DOM time calibration is refined exploiting the signals from the LED beacons on each DOM and using atmospheric muons \cite{ICRC_calib}. This paper does not cover the inter-DOM calibration as the reported results are obtained from standalone DOM data.

\section{Data sample} 
\label{s:sample}

The selected data taking period for the ARCA2 detector ranges from December 23, 2016 to March 2, 2017. At the beginning of February 2017, an interruption of the operation of ARCA2 of approximately one week happened due to an electrical problem in the onshore infrastructure. 
Four DOMs of ARCA2 were not active during the data taking period used in this analysis.
The selected data taking period for the ORCA1 detector is from November 9, 2017 to December 13, 2017. From November 23 2017, the timeslice data of ORCA 1 has been downscaled by a factor 20, in order to reduce the amount of data written on disk. 
In this analysis, the data from a DOM with one or more PMTs in high rate veto (see Section \ref{s:detector}) are ignored on a timeslice basis. The data arriving from the rest of the detector are kept.
The average fraction of PMTs in high rate veto  is of a few per mil in ARCA and a few per cent in ORCA.

\begin{figure}
\begin{center}
\includegraphics[width=0.75\textwidth]{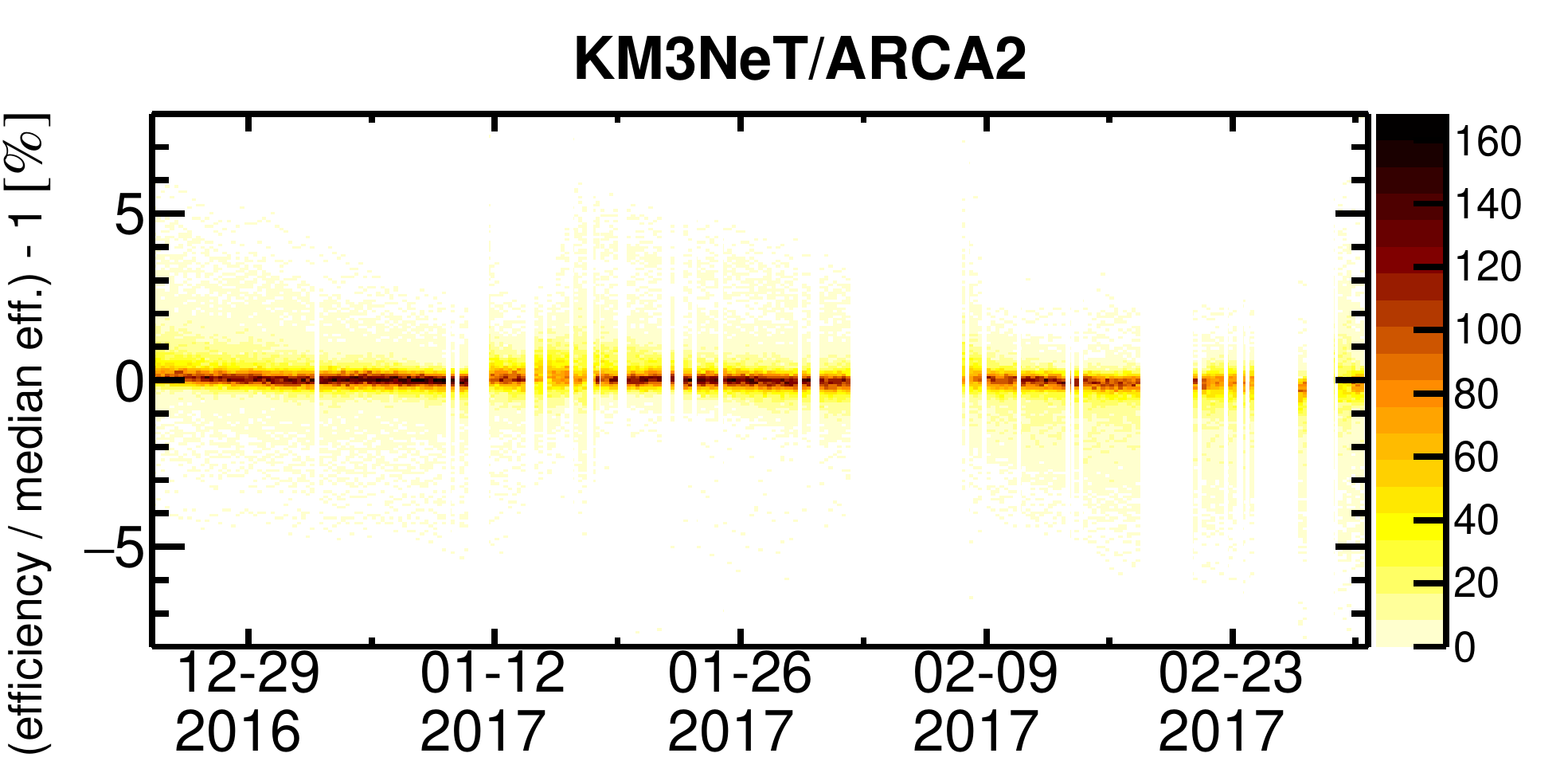}\\
\includegraphics[width=0.75\textwidth]{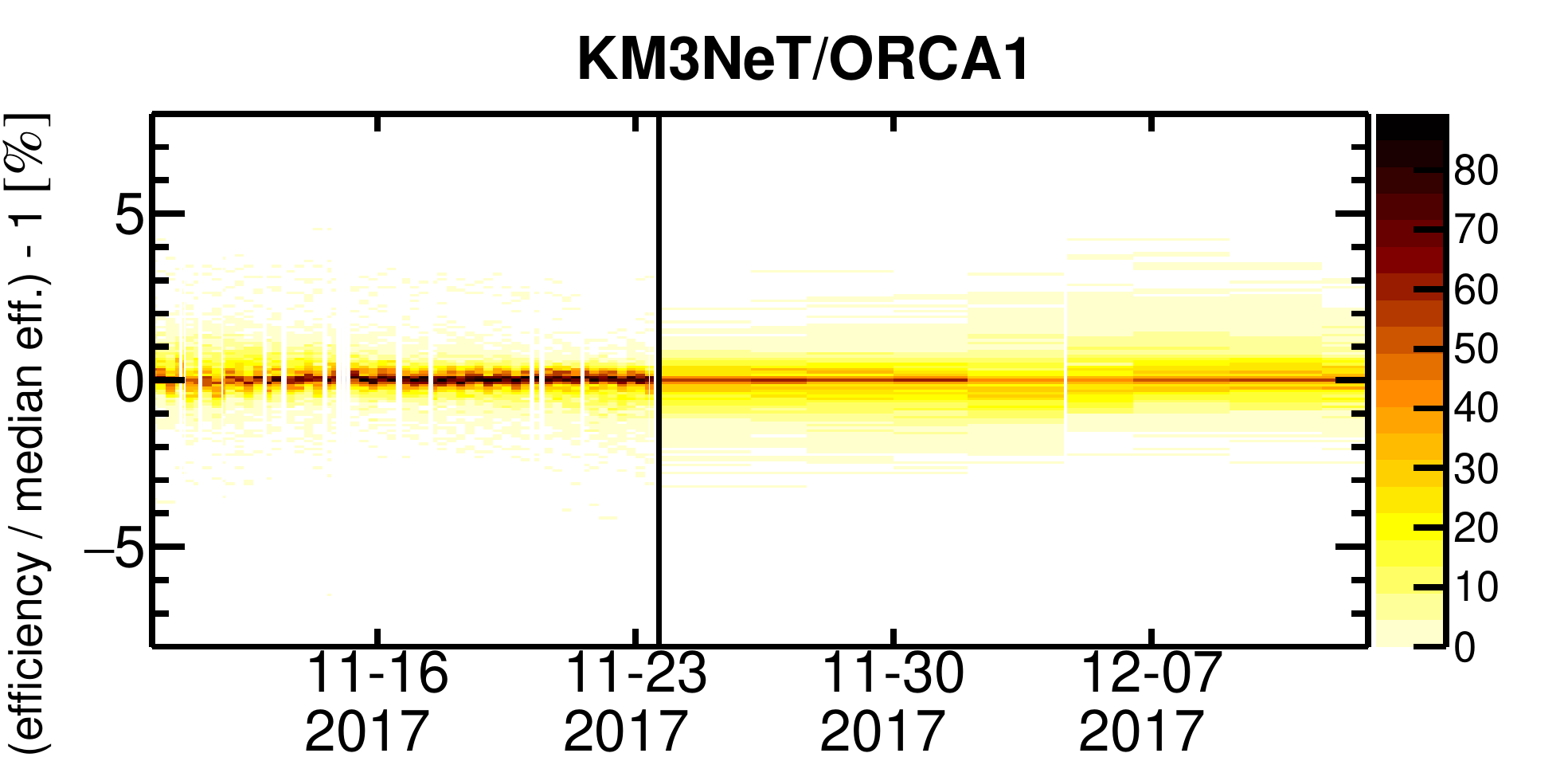}\\
\caption{Deviation of the estimated photon detection efficiency for each PMT with respect to its global median efficiency for the ARCA2 and ORCA1 detectors as a function of time. The color scale indicates the number of PMTs in each bin. Vertical white bands reflect the periods without data-taking; the vertical black line represents the time at which  the L1 data stream downscaling was introduced (see text).}
\label{fig:stability}
\end{center}
\end{figure}

The combined PMT time offset and PMT photon detection efficiency calibration has been performed over segments of 6 hours of data taking, which provide sufficient statistics to prevent fluctuations in the parameter estimations used for data analysis and simulation. 
In Figure \ref{fig:stability}, the stability of photon detection efficiencies is evaluated by plotting for each individual PMT the deviation of the efficiency with respect to its median efficiency estimated from the entire period of data taking. 
The efficiencies are stable in time, as the root mean square (RMS) of the distribution of the  deviations is 0.5\% for ORCA1 and 0.8\% for ARCA2 in the considered data taking period. The fitted PMT photon detection efficiencies of the six lowest DOMs of ORCA1 are approximately 15\% lower than the average efficiency of the other ORCA and ARCA DOMs. A similar pattern is observed in the single hit rates of these PMTs. Investigations are in progress to understand the origin of this difference.

\section{Monte Carlo simulations} 
\label{s:mc}

The response of the detector has been studied in detail using Monte Carlo simulated atmospheric muon events.
The Monte Carlo chain is based on a multi-stage approach comprising an event generator, a simulator of the Cherenkov light and a stage which covers the combination of the simulation of the PMT response, the readout and the onshore data filtering. 

Atmospheric muon events have been generated using the MUPAGE package \cite{MUPAGE}. This software provides a parametric calculation of the underwater muon flux of atmospheric muon bundles, based on a full Monte Carlo simulation of the primary cosmic ray interactions and shower propagation in the atmosphere. With MUPAGE, muon bundles have been generated with a bundle energy $E_b$ above $\SI{10}{GeV}$ over the surface of a cylinder which extends the instrumented volume by \SI{280}{m}. Light emitted by muons simulated outside of this volume have a negligible probability to produce at least one L1 coincidence.

The tracking of muons in seawater and subsequent production and propagation of the Cherenkov light are implemented in the KM3 \cite{km3} software package. KM3 uses tabulated results from a full GEANT3.21 \cite{geant} simulation of relativistic muons and electromagnetic showers. In this, the Cherenkov light production, propagation, scattering and absorption in seawater are taken into account. In addition, KM3 also accounts for the effects associated with the DOM structure, such as the reflector rings and the light absorption in the glass and optical gel.

Using custom KM3NeT software, the detector response is simulated. Random optical background hits are added according to single hit and coincidence rates observed in data. The PMT response is then simulated according to the measured transit time distribution \cite{PMT} and PMT parameters (photon detection efficiency, gain and gain-spread). In this, the PMT photon detection efficiencies can either be set to the nominal value, or to the measured values using the calibration procedure described in Section \ref{s:calib}. The software is designed to produce a data format identical to the one produced by the DAQ system. As a result, the same algorithms can be applied independently on data and simulations.
\section{Depth dependence of atmospheric muon flux} 
\label{s:muons}

\subsection{Optical background discrimination}

The multi-PMT design of the DOM provides information on the number of photons as well as on their arrival times and directions. This information is used to discriminate between the signals from atmospheric muons, bioluminescence and ${}^{40}$K backgrounds.
The discriminating power is mainly achieved by exploiting coincidences between different PMTs in the same DOM. In this analysis, a coincidence is defined as a sequence of hits in a DOM within a time window of $\SI{15}{ns}$. The number of hit PMTs in a coincidence is called \textit{multiplicity}. 

\begin{figure}
\begin{center}
\includegraphics[width=0.75\textwidth]{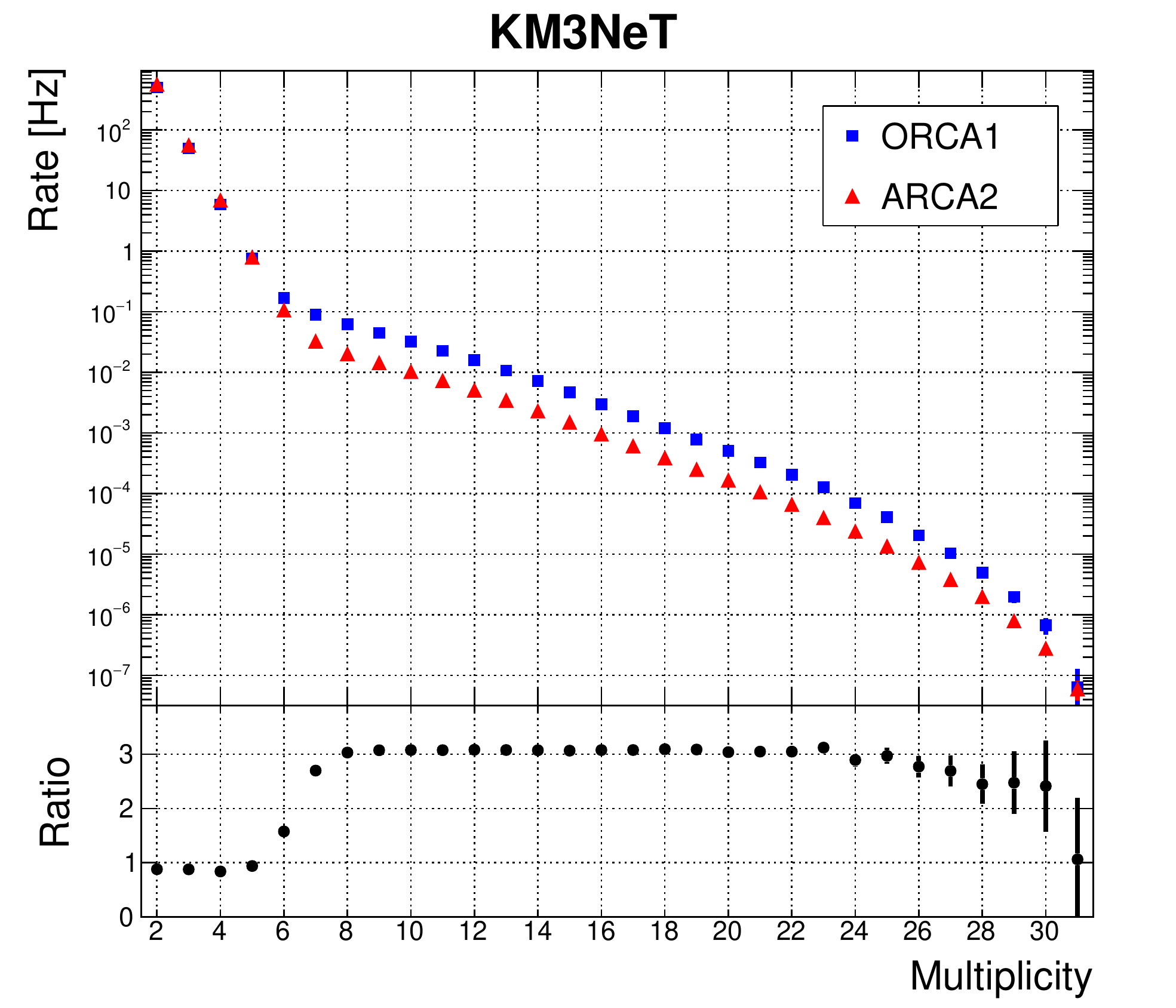}
\caption{Top: coincidence rates as a function of the multiplicity for the ORCA1 and ARCA2 detectors averaged over all the DOMs of each detector. Bottom: ratio between ORCA1 and ARCA2 coincidence rates. Up to a multiplicity of six, the coincidence rate is dominated by $^{40}$K decays. Above a multiplicity of seven, atmospheric muons dominate. Only statistical errors are shown.}
\label{fig:multiplicity_rates}
\end{center}
\end{figure}

In Figure \ref{fig:multiplicity_rates}, the coincidence rates, averaged over the DOMs of the ARCA2 and ORCA1 detectors respectively, are shown as a function of the multiplicity. Only statistical uncertainties are indicated. The random combinatorial background has been subtracted and the times of the hits have been calibrated as described in Section~\ref{s:calib}.

The contribution from $^{40}$K decays is dominant up to a multiplicity of seven \cite{ppm-du}. Conductivity measurements indicate the salinity in seawater to be independent of depth and detector site \cite{orca_site,arca_site}. 
Due to the difference in the average efficiency between the optical modules of the two detectors (see Section \ref{s:sample}), the $^{40}$K coincidence rates observed in ORCA1 are lower than the rates observed in ARCA2. This difference is about 12\% at multiplicity 2, which is consistent with a quadratic dependence of the rate on the average PMT efficiency.

\begin{figure}
\begin{center}
\includegraphics[width=0.75\textwidth]{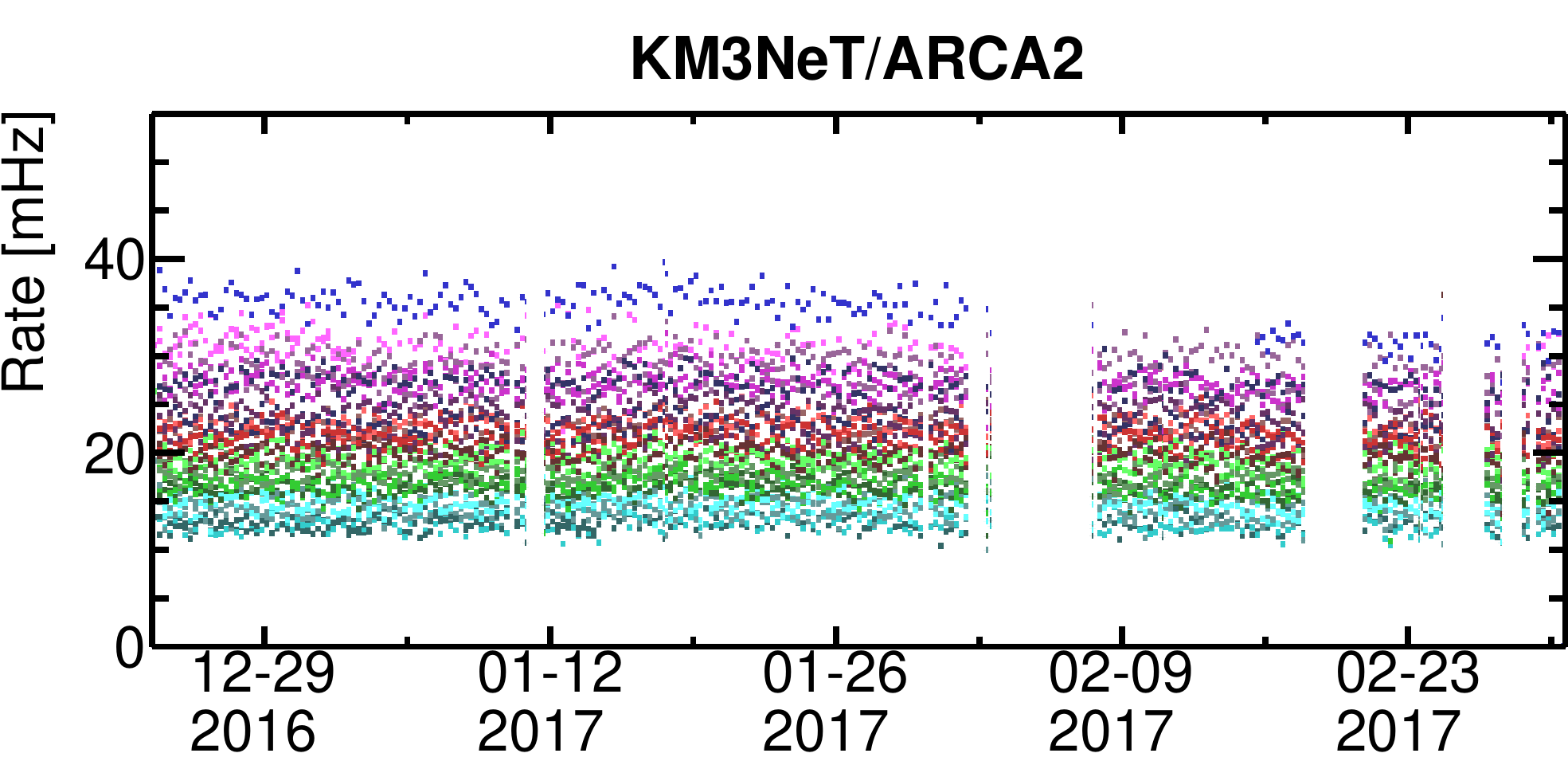} \\ 
\includegraphics[width=0.75\textwidth]{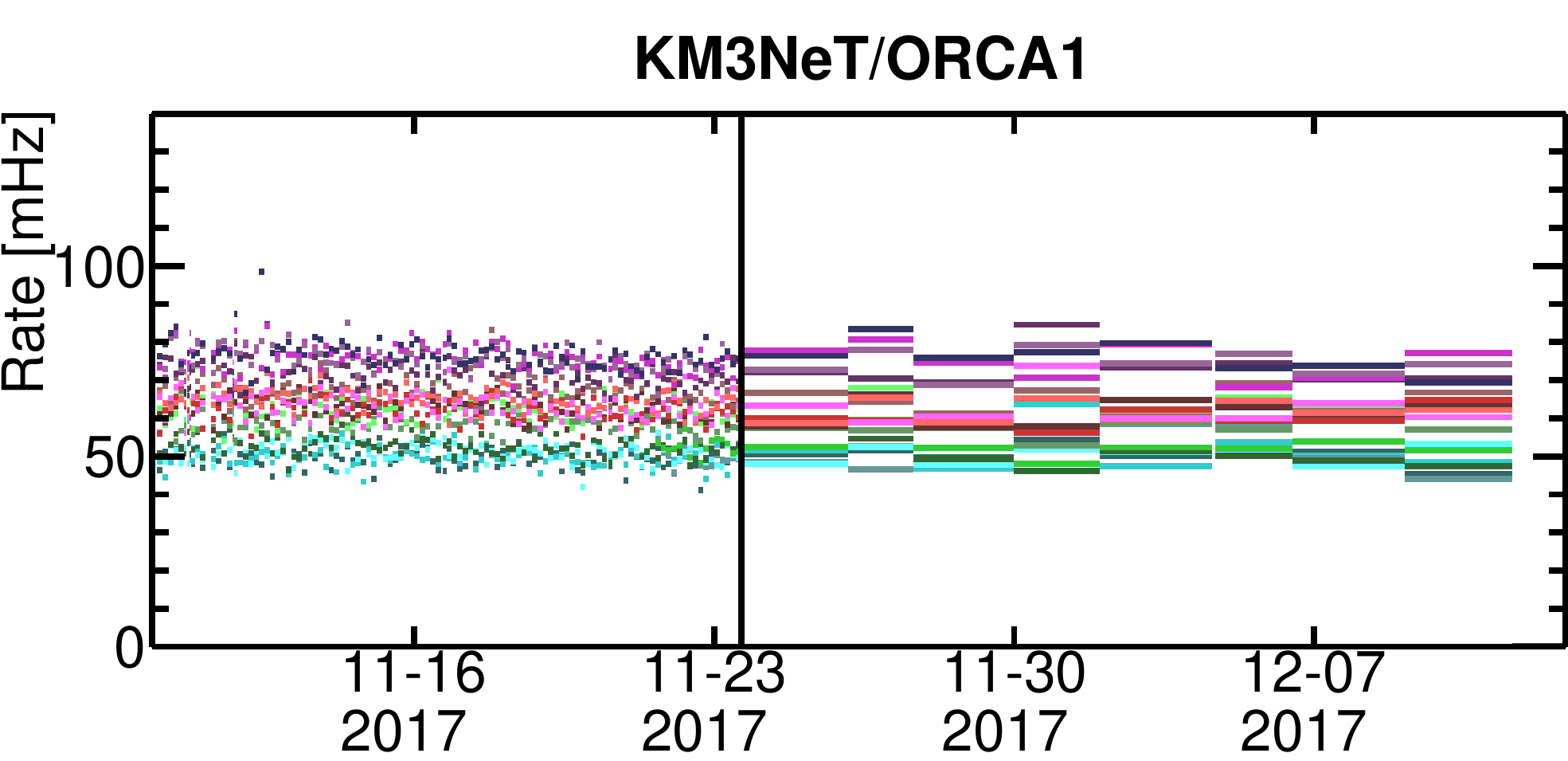}
\caption{Rate of  multiplicity $\geq 8$ coincidences as a function of time for the DOMs of the ARCA2 and ORCA1 detectors. Each point corresponds to one run and every colour to a single DOM. Vertical white bands are periods without data-taking. The vertical black line in the bottom plot represents the date at which the L1 data stream is downscaled (see Section \ref{s:sample}).}
\label{fig:ratestability}
\end{center}
\end{figure}

Coincidences from atmospheric muons dominate at a multiplicity of eight and higher as muons and muon bundles can potentially illuminate all the PMTs of a single DOM.
The ratio between ARCA2 and ORCA1 at high multiplicities shows a factor three difference due to the different average depths of the DOMs, which is around \SI{2310}{m} for ORCA and \SI{3070}{m} for ARCA. In Figure \ref{fig:ratestability}, the stability of the coincidences rates in the multiplicity region dominated by atmospheric muons is shown. 
Four ARCA2 DOMs lost one PMT during the considered data taking, resulting in a decrease of the respective rates. The data from the affected DOMs are rejected in the following analysis when evaluating the muon flux.

\begin{figure}
\begin{center}
\includegraphics[width=0.49\textwidth]{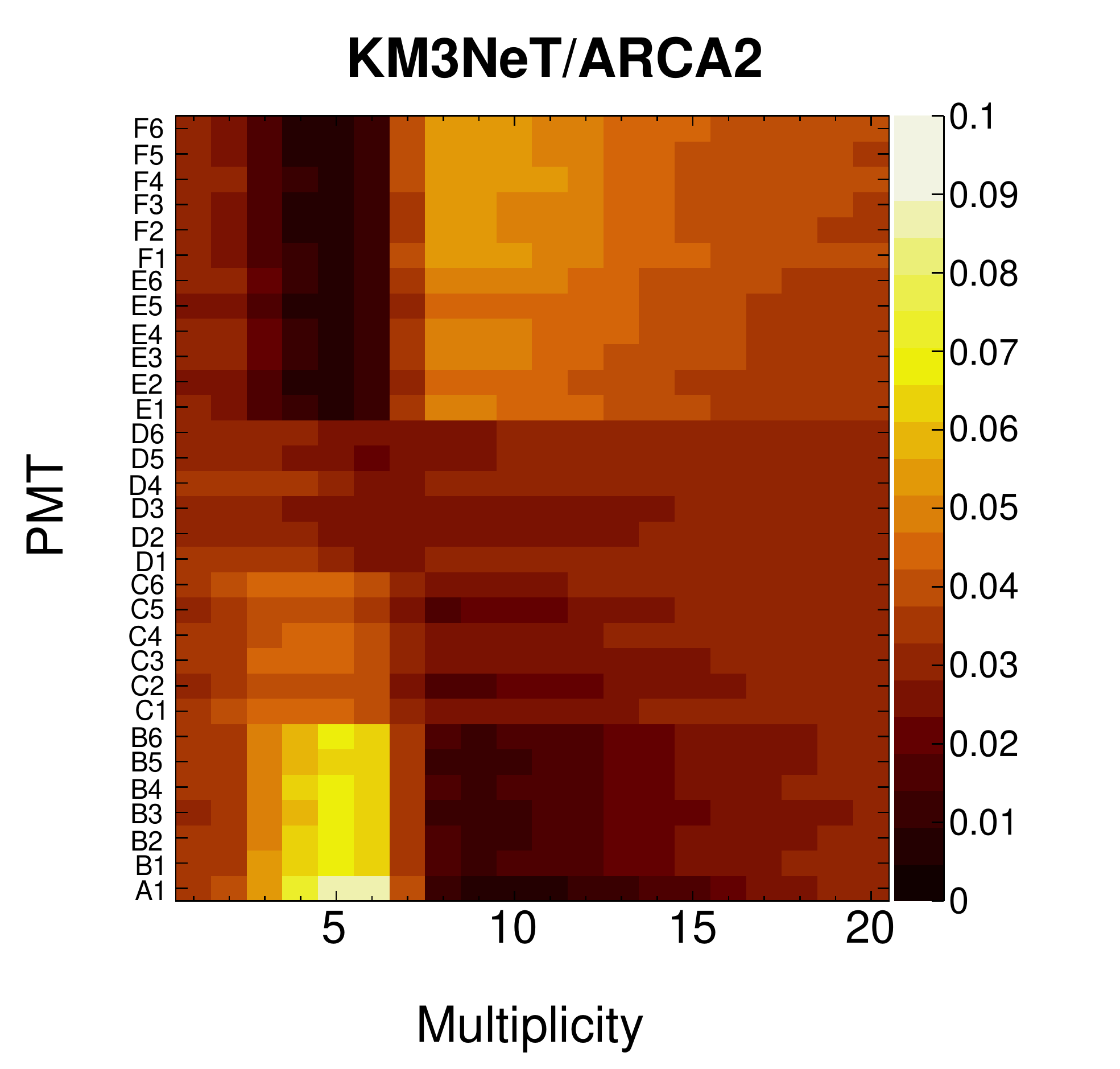} \ \ 
\includegraphics[width=0.49\textwidth]{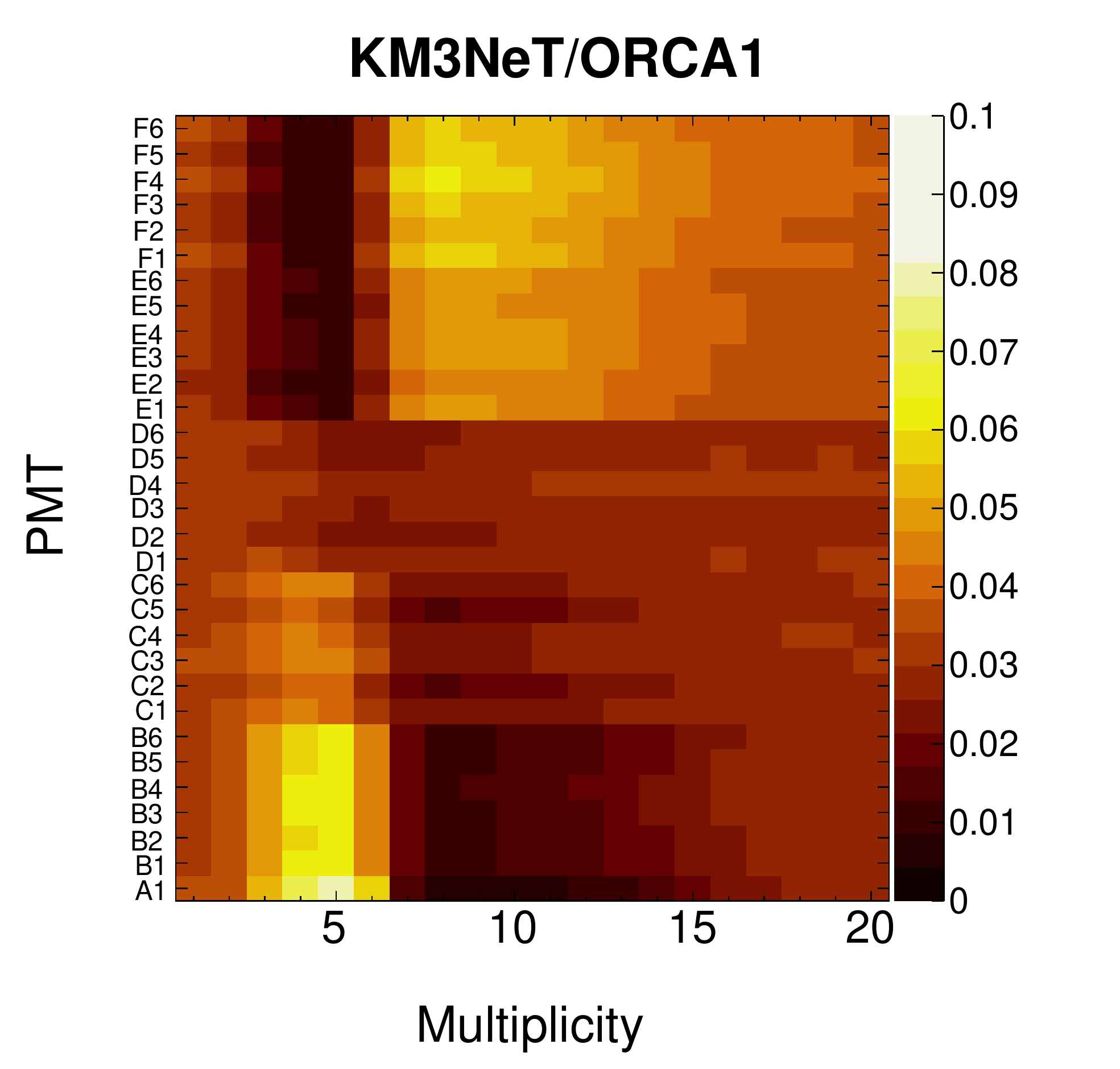}
\caption{Probability density function of each PMT contribution to coincidence rates as a function of the multiplicity (each abscissa bin is normalised to unity).  The PMT is identified by the ring (letter) and the position of the PMT on the ring (number). The first address (A1) refers to the vertical down-facing PMT, rings from B to D belong to the lower hemisphere, while rings E and F belong to the upper hemisphere of the DOM (see Figure~\ref{f:DOMpicture}). Above a multiplicity of 20, statistical fluctuations dominate the pattern, and are therefore left out of the figure.}
\label{fig:occupancy}
\end{center}
\end{figure}

The relative contribution of PMTs to coincidences as a function of the multiplicity is shown in Figure \ref{fig:occupancy}. The lower DOM hemisphere is more populated at lower multiplicities due to the higher number of close-by PMTs, resulting in more coincidences from $^{40}$K decays. Shadowing effects of the rope-mounting structures can be observed for PMTs C2 and C5. On the other hand, the contribution to high multiplicities comes mostly from the upper hemisphere. This reflects the downgoing direction of atmospheric muons.

\subsection{Muon-induced coincidence rates} \label{ss:results}

\begin{figure}
\begin{center}
\includegraphics[width=0.75\textwidth]{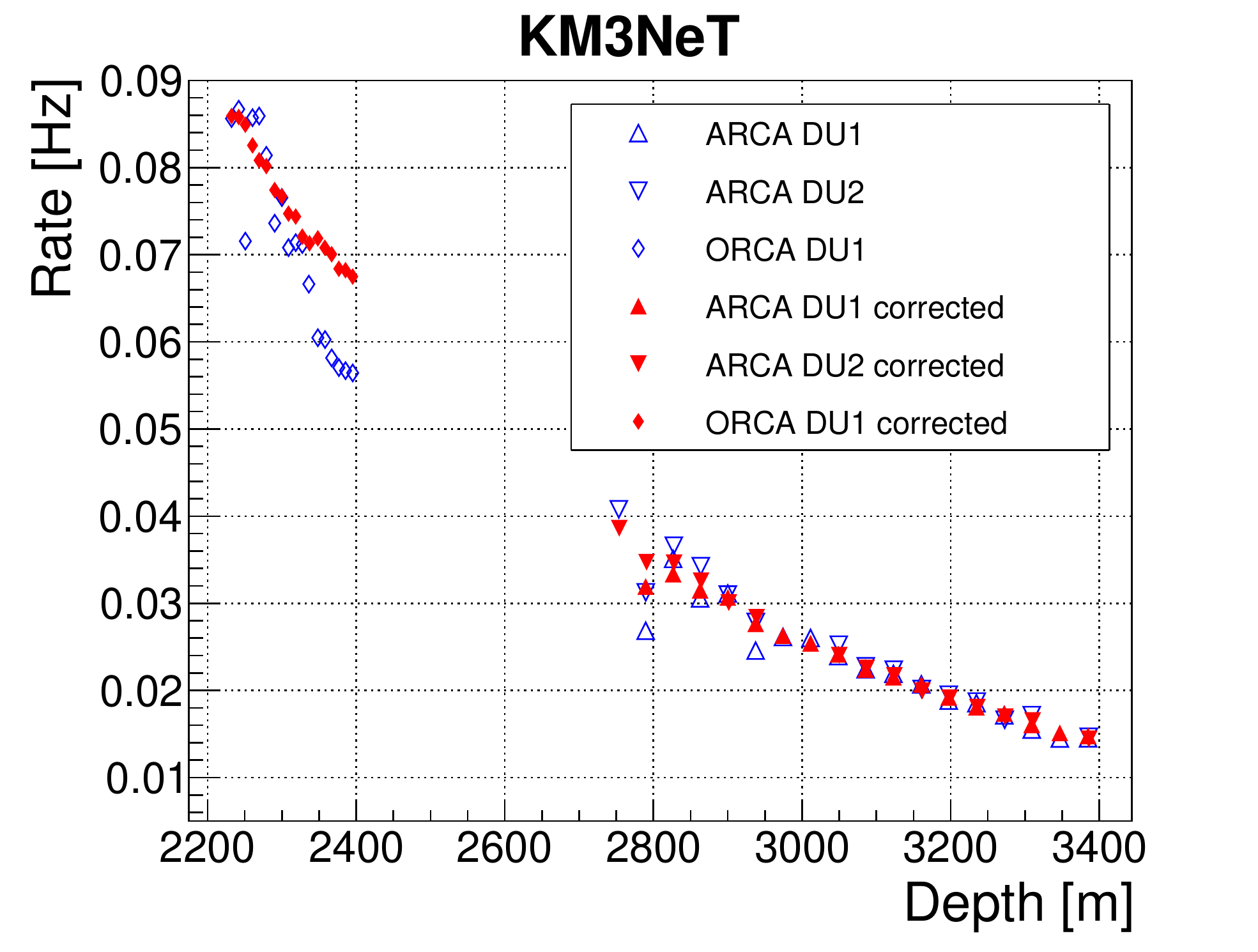}
\caption{Multiplicity $\geq8$ coincidence rate of all DOMs as function of depth below the sea level. The coincidence rates for the ARCA2 and ORCA1 detectors are reported as measured (blue points) and after the correction for the PMT photon detection efficiencies (red points).  Statistical uncertainties are included and smaller than markers.}
\label{fig:depthdependence}
\end{center}
\end{figure}

As a result of energy losses in seawater, a lower rate of atmospheric muons is observed at larger depths. This is reflected in the average coincidence rates for multiplicities $\geq8$ between the two detectors (Figure \ref{fig:multiplicity_rates}). The $\geq 8$ multiplicity coincidence rate for all the active DOMs of the two detectors is shown in Figure \ref{fig:depthdependence} as a function of the depth of each DOM.

Differences in the PMT photon detection efficiencies between the DOMs affect the measured rates, thereby affecting also the depth-dependence relation. In order to correct for this, two Monte Carlo simulations have been performed. %In the first set of simulations, referred to as `uniform', the photon detection efficiencies are set to the nominal value in order to establish the MC normalisation. 
In the first set of simulations, referred to as `uniform', the photon detection efficiencies are set to the average efficiency obtained with the calibration procedure for a set of typical DOMs, in order to establish the MC normalisation.
In the second, the photon detection efficiencies are set to their measured values (see Section \ref{s:calib}). The result is referred to as `calibrated'. 

The ratio between the simulated rates is used to correct the measured rates for each DOM, $R^{\textrm{data}}_{\textrm{measured}}$, applying the formula:

\begin{equation} \label{eq:correction}
 R^{\textrm{data}}_{\textrm{corrected}} = R^{\textrm{data}}_{\textrm{measured}} \cdot \frac{R^{\textrm{MC}}_{\textrm{uniform}}}{R^{\textrm{MC}}_{\textrm{calibrated}}} \: .
\end{equation}
In Figure \ref{fig:depthdependence}, the efficiency-corrected coincidence rates are shown in red. For the lowest six DOMs of ORCA1, the correction factor is larger than for ARCA2. This can be explained by the lower efficiencies of the PMTs of ORCA1 (see Section \ref{s:sample}).

\subsection{Determination of the atmospheric muon flux} \label{ss:model}

\begin{figure*}
\begin{center}
\includegraphics[width=0.8\textwidth]{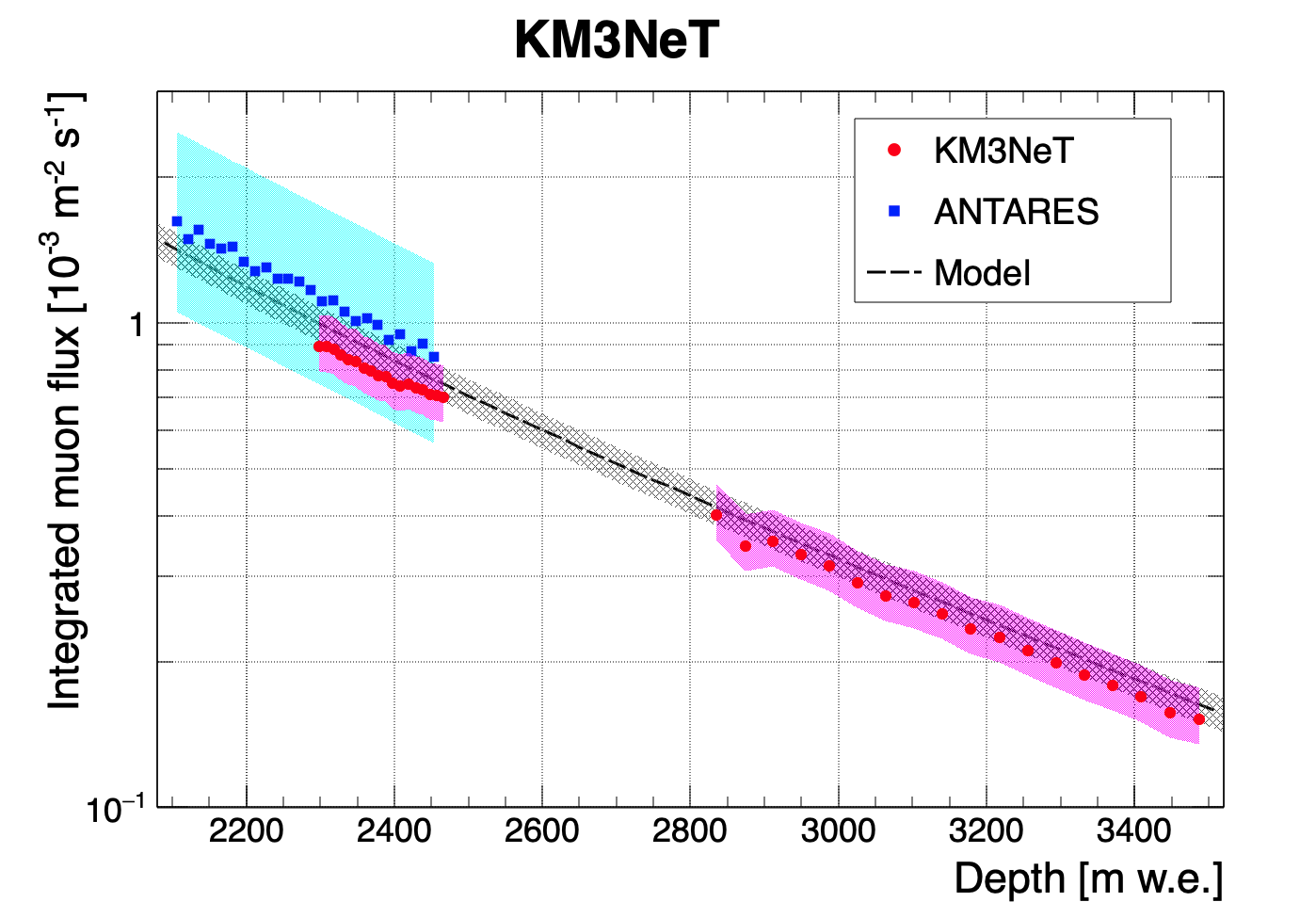}
\caption{Integrated atmospheric muon flux measured with the ORCA1 and ARCA2 detectors as a function of depth below the sea level (red points). The systematic errors are displayed as light red shadowed areas. The Bugaev model of the atmospheric muon flux is drawn with a dashed black line (quoted errors are the grey shadowed area), see text for model description. ANTARES data from \cite{zaborov_antares} are included as blue points for comparison (systematic errors are the light blue shadowed area). The depth is expressed in water equivalent (w.e.). Statistical uncertainties are included and smaller than markers.}
\label{fig:depthdependence_avg}
\end{center}
\end{figure*}

The efficiency-corrected rates are used in the following to measure the atmospheric muon flux. The single DOM response to atmospheric muons in terms of multiplicity $\geq8$ coincidences is quantified in terms of \emph{effective area}. A dedicated MUPAGE simulation, using an up-to-date PMT model reflecting our best knowledge of the angular acceptance and quantum efficiency is used to estimate the coincidence rates. The effective area ($A_{eff}$) is the ratio between the coincidence rate measured by a simulated DOM and the muon flux estimated at the boundary surface of the generation volume at the same depth. From the simulation, the average value of the effective area for multiplicity $\geq 8$ coincidences is $96^{+5}_{-12}\SI{}{m^{2}}$. The uncertainty taken into account here will be discussed in Section \ref{ss:systematics}. The muon flux measurement is obtained by the ratio between the corrected measured coincidence rates and the DOM effective area. The flux as a function of the depth expressed in metres of water equivalent is shown in Figure \ref{fig:depthdependence_avg}. As a comparison, the model from Bugaev et al. \cite{Bugaev} is shown, together with the ANTARES measurement previously presented in \cite{zaborov_antares}.

The angle-integrated flux as a function of depth for the considered model has been calculated from the formulae provided in Ref. \cite{Klimushin:2000cy}. In order to compare with the data, the flux model is here described with a simple analytic expression in the form of a vertical flux $I_\mu (d, \theta = 0)$ corrected with a factor $C(d)$ that accounts for the angular integration.
In the depth range of interest, the resulting expression reads as follows (where the water equivalent depth $d$ accounts for the density of seawater):

\begin{equation}
 I_\mu (d) = \frac{I_\mu(d, \theta = 0)}{C(d)} = \frac{A_1 \cdot e^{A_2 \cdot d} + A_3 \cdot e^{A_4 \cdot d}}{B_1 + B_2 \cdot d} \: ,
\end{equation}

\[ A_1 = \SI{1.31e-5}{cm^{-2} s^{-1} sr^{-1}} \: , \; A_2 = \SI{-2.91e-3}{m^{-1}} \: , \]
\[ A_3 = \SI{7.31e-7}{cm^{-2} s^{-1} sr^{-1}} \: , \; A_4 = \SI{-1.17e-3}{m^{-1}} \: , \]
\[ B_1 = \SI{4.16e-1}{sr^{-1}} \: , \; B_2 = \SI{1.07e-4}{m^{-1}sr^{-1}} \: . \]

\noindent In this, the $A_i$ parameters define the depth dependence of the vertical flux and the $B_j$ parameters the integration factor. This parameterisation is valid for a flux of muons with energies above 1 GeV. 
From \cite{Klimushin:2000cy}, a conservative error of $\pm 8$\% is assumed on the parameterisation of the muon flux.

ANTARES and KM3NeT data are compatible with the model within the systematic uncertainties.
If the normalisation of the model is fitted to the data, the RMS of the differences between model and data is lower than 2\%.

\subsection{Evaluation of systematic uncertainties} \label{ss:systematics}

The systematic error assigned to the muon flux measurement is estimated from the uncertainties on the determined PMT efficiencies, the knowledge of absorption length of light in water and the DOM effective area used to calculate the muon flux from the $\geq8$ multiplicity coincidence rate.

The uncertainty on the absolute PMT efficiency is estimated from the comparison between the data and the nominal DOM simulation introduced in Section~\ref{s:calib}.
The mean difference of the coincidence rates across all PMT pairs averages at 5\%, including the effect of the Gaussian modeling of the $^{40}$K hit time difference distribution (Figure~\ref{f:double_coinc}). This value is taken as the  systematic uncertainty on the PMT efficiency normalisation. From calibration studies, the uncertainty on the relative efficiency of individual PMTs is assigned a systematic uncertainty at the 5\% level. Conservatively, this error is propagated to the overall DOM efficiency normalisation and added in quadrature to the uncertainty on the absolute PMT efficiency. The total effect of the uncertainties on the PMT efficiency applied to the DOM coincidence rates is then of 7\%.

The effect of the estimated 10\% uncertainty on the light absorption length in water properties has been studied with a dedicated simulation. The corresponding systematic uncertainty on the measurement of the muon flux is 6\%.

The effective area is expected to be dependent on the angular distribution and on the average bundle multiplicity of atmospheric muons. In turn, the two factors depend on depth. The bundle multiplicity is found to be the dominant variable in the determination of the effective area. For the measurement of the muon flux, the effective area is assumed to be constant with an uncertainty of $^{+5\%}_{-13\%}$. This estimation covers the effect of the bundle multiplicity variation with depth and the uncertainty on its absolute normalisation. The latter is assigned asymmetrically, as MUPAGE is currently believed to overestimate the bundle multiplicity.

The total systematic uncertainty on the muon flux measurement obtained by the sum in quadrature of the evaluated factors amounts to $^{+16\%}_{-11\%}$. 
Since the measured flux is inversely proportional to the effective area, the sign of the asymmetry is reversed.

\section{Conclusions} 
\label{s:conclusion}

Data from the first three detection units of the KM3NeT ARCA and ORCA detectors have been used to measure the atmospheric muon flux over a wide range of depths. The coincidence rate of multiplicity $\geq 8$ is used to select a high-purity sample of atmospheric muons on a DOM-by-DOM basis. With this measurement, the calibration procedure based on coincidence hits from ${}^{40}$K decays is verified. The estimated photon detection efficiencies and time offsets of the PMTs of the KM3NeT detectors are shown to be stable over time. 
The measured atmospheric muon flux is found to be compatible with the Bugaev flux parameterisation over the entire depth range considered, extending the previous measurement performed by ANTARES. This approach provides a precise estimation of the total muon flux along the detector depth, complementing studies of atmospheric muons based on track reconstruction. Independently of the physical site and of the depth of the detector, the results of KM3NeT ORCA1 and ARCA2 are in agreement with the expected atmospheric muon flux over a range of more than one kilometre.

\section{Acknowledgements}
The authors acknowledge the financial support of the funding agencies:
Agence Nationale de la Recherche (contract ANR-15-CE31-0020),
Centre National de la Recherche Scientifique (CNRS), 
Commission Europ\'eenne (FEDER fund and Marie Curie Program),
Institut Universitaire de France (IUF),
IdEx program and UnivEarthS Labex program at Sorbonne Paris Cit\'e (ANR-10-LABX-0023 and ANR-11-IDEX-0005-02),
Paris \^Ile-de-France Region,
France;
Shota Rustaveli National Science Foundation of Georgia (SRNSFG, FR-18-1268),
Georgia;
Deutsche Forschungsgemeinschaft (DFG),
Germany;
The General Secretariat of Research and Technology (GSRT),
Greece;
Istituto Nazionale di Fisica Nucleare (INFN),
Ministero dell'Istruzione, dell'Universit\`a e della Ricerca (MIUR),
PRIN 2017 program (Grant NAT-NET 2017W4HA7S)
Italy;
Ministry of Higher Education, Scientific Research and Professional Training,
Morocco;
Nederlandse organisatie voor Wetenschappelijk Onderzoek (NWO),
the Netherlands;
The National Science Centre, Poland (2015/18/E/ST2/00758);
National Authority for Scientific Research (ANCS),
Romania;
Plan Estatal de Investigaci\'on (refs.\ FPA2015-65150-C3-1-P, -2-P and -3-P, (MINECO/FEDER)), Severo Ochoa Centre of Excellence program (MINECO), Red Consolider MultiDark, (ref. FPA2017-90566-REDC, MINECO) and Prometeo and Grisol\'ia programs (Generalitat Valenciana),
``la Caixa'' Foundation (ID 100010434) through the fellowship LCF/BQ/IN17/11620019, and the European Union's Horizon 2020 research and innovation programme under the Marie Sk\l{}odowska-Curie grant agreement no. 713673,
Spain.

\end{document}